\PassOptionsToClass{10pt}{revtex4-1}
\documentclass[aps,prl,showpacs,floatfix,twocolumn,byrevtex,superscriptaddress]{revtex4-1}

\usepackage{amsmath}
\usepackage{amssymb}
\usepackage{amstext}
\usepackage{amsopn}
\usepackage{amsfonts}
\usepackage{amsxtra}
\usepackage[english]{babel}
\usepackage{graphicx}
\usepackage{bm}
\usepackage{multirow}
\usepackage{dcolumn}
\usepackage{color}
\usepackage{textgreek}
\usepackage{hyperref}
\usepackage{todonotes}
\usepackage{verbatim}
\usepackage{soul}

\def\mathbi#1{\ensuremath{\textbf{\em #1}}}
\def\Q{\mathbi{Q}}
\begin{document}

\title{Phononic Helical Nodal Lines with $\mathcal{PT}$ Protection in MoB$_{2}$}

\author{T. T. Zhang}
\affiliation{Institute of Physics, Chinese Academy of Sciences, Beijing 100190, China}
\affiliation{University of Chinese Academy of Sciences, Beijing 100049, China}
\affiliation{Department of Physics, Tokyo Institute of Technology, Ookayama, Meguro-ku, Tokyo 152-8551, Japan}
\affiliation{Tokodai Institute for Element Strategy, Tokyo Institute of Technology, Nagatsuta, Midori-ku, Yokohama, Kanagawa 226-8503, Japan}
\author{H. Miao}\email[]{hmiao@bnl.gov}
\affiliation{Condensed Matter Physics and Materials Science Department, Brookhaven National Laboratory, Upton, New York 11973, USA}
\author{Q. Wang}
\affiliation{Department of Physics and Beijing Key Laboratory of Opto-Electronic Functional Materials and Micro-devices, Renmin University of China, Beijing, China}
\author{J. Q. Lin}
\affiliation{Condensed Matter Physics and Materials Science Department, Brookhaven National Laboratory, Upton, New York 11973, USA}
\affiliation{University of Chinese Academy of Sciences, Beijing 100049, China}
\affiliation{School of Physical Science and Technology, ShanghaiTech University, Shanghai 201210, China}
\author{Y. Cao}
\affiliation{Materials Science Division, Argonne National Laboratory, Argonne, IL 60439, USA}
\author{G. Fabbris}
\affiliation{Advanced Photon Source, Argonne National Laboratory, Argonne, Illinois 60439, USA}
\author{A. H. Said}
\affiliation{Advanced Photon Source, Argonne National Laboratory, Argonne, Illinois 60439, USA}
\author{X. Liu}
\affiliation{School of Physical Science and Technology, ShanghaiTech University, Shanghai 201210, China}
\author{H. C. Lei}\email[]{hlei@ruc.edu.cn}
\affiliation{Department of Physics and Beijing Key Laboratory of Opto-Electronic Functional Materials and Micro-devices, Renmin University of China, Beijing, China}
\author{Z. Fang}
\affiliation{Institute of Physics, Chinese Academy of Sciences, Beijing 100190, China}
\affiliation{Songshan Lake Materials Laboratory, Dongguan, Guangdong 523808, China}
\author{H. M. Weng}\email[]{hmweng@iphy.ac.cn}
\affiliation{Institute of Physics, Chinese Academy of Sciences, Beijing 100190, China}
\affiliation{Songshan Lake Materials Laboratory, Dongguan, Guangdong 523808, China}
\author{M. P. M. Dean}\email[]{mdean@bnl.gov}
\affiliation{Condensed Matter Physics and Materials Science Department, Brookhaven National Laboratory, Upton, New York 11973, USA}

\begin{abstract}

While condensed matter systems host both Fermionic and Bosonic quasi-particles, reliably predicting and empirically verifying topological states is only mature for Fermionic electronic structures, leaving topological Bosonic excitations sporadically explored. This is unfortunate, as Bosonic systems such a phonons offer the opportunity to assess spinless band structures where nodal lines can be realized without invoking special additional symetries to protect against spin-orbit coupling. Here we combine first-principles calculations and meV-resolution inelastic x-ray scattering to demonstrate the first realization of parity-time reversal ($\mathcal{PT}$) symmetry protected helical nodal lines in the phonon spectrum of MoB$_{2}$. This structure is unique to phononic systems as the spin-orbit coupling present in electronic systems tends to lift the degeneracy away from high-symmetry locations. Our study establishes a protocol to accurately identify topological Bosonic excitations, opening a new route to explore exotic topological states in crystalline materials.

\end{abstract}
\maketitle
 
\begin{figure}
\includegraphics[scale=1]{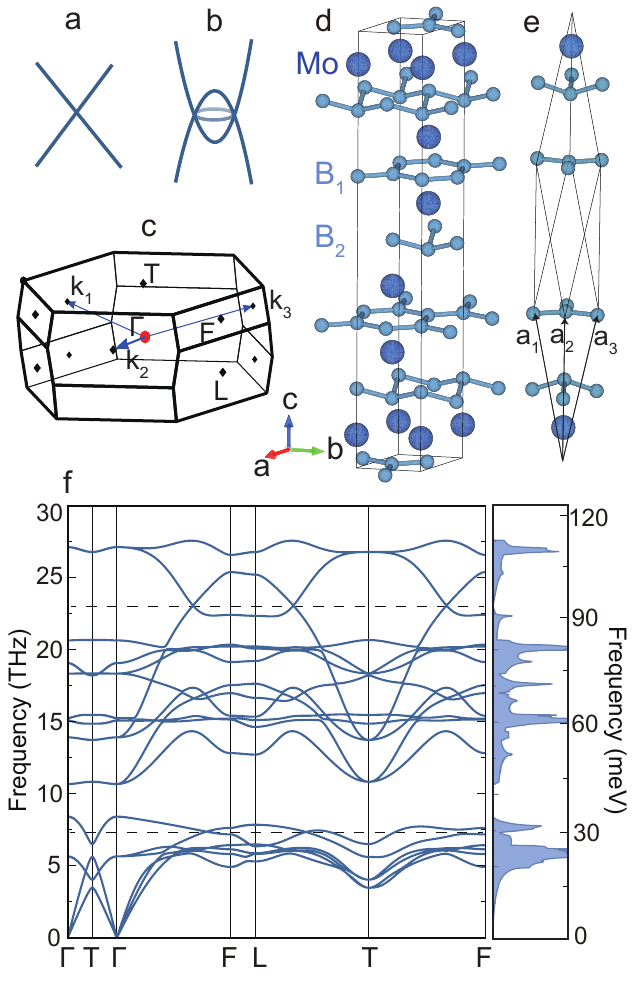}\caption{Schematics of (a) Weyl point with $z$-invariance and (b) topological nodal line with $z_2$-invariance. The  structure in the center of (b) (colored in light blue) represents the extended nature of the crossing, which traces out a loop in reciprocal space. (c) is the BZ of MoB$_{2}$, where $k_{1}$, $k_{2}$ and $k_{3}$ are the reciprocal basis of the primitive cell. (d) and (e) are the unit cell and primitive cell of MoB$_{2}$ respectively. \textit{a}, \textit{b} and \textit{c} specify the conventional unit cell directions, while $\textit{a}_1$, $\textit{a}_2$ and $\textit{a}_3$ are base vectors of the primitive cell. The high-symmetry points, $\Gamma$:(0,0,0), T:(0.5,0.5,0.5), F:(0,0.5,-0.5) and L:(0.5,1,0), are defined in the primitive cell notation. (f) is the DFPT calculated phonon spectra and the phonon DOS of MoB$_{2}$.} 
\label{fig:FIG1}
\end{figure}

The discovery of topological insulators and various topological semimetals in electronic band structures has ignited extensive research to classify various topological states and identify materials whose electronic structures are topologically non-trivial. A recent breakthrough is the development of symmetry-based indicator theories  \cite{po2017symmetry,song2018quantitative,song2018diagnosis} and compatibility relations \cite{bradlyn2017TQC}, which have been used to thoroughly diagnose topological states in the electronic structures of most known materials. Despite this transformative progress on electronic Fermionic quasi-particles, the systematic study of topological Bosonic states in crystalline materials is still in its infancy. In particular, it is unclear how to realize new topological Bosonic excitations. 

Phonons are the most basic emergent Boson of crystalline lattices, describing the collective motion of the underlying atoms. They strongly affect the physical properties of crystalline materials and play a pivotal role in various electronic phases, such as conventional superconductivity and charge density waves. Phonons in crystalline materials are different from those in mechanical metamaterials, where the former are true quantum objects and correlate with electronic and magnetic exciations while the later are derived from classical equations \cite{Wang2015,Stenull2016}. Conceptually, phonon band dispersions share similar symmetry properties to spinless electronic band structures \cite{Top_classifacation_symmetry}, and hence can host topologically nontrivial band crossings, such as Weyl points \cite{miao2018observation,zhang2018double} and Dirac points \cite{li2017dirac,yao2018topological}. In these cases, the Weyl points are the sources of nonzero Berry flux in momentum space with topological $z$-invariance \cite{fang2003anomalous,armitage2018weyl}. However, when both time-reversal symmetry, $\mathcal{T}$, and inversion symmetry, $\mathcal{P}$, are present, an isolated Weyl-point, as shown in Fig.~\ref{fig:FIG1}a, no longer exists as the $\mathcal{PT}$ operation restricts the Berry curvature to be zero. In cases where $\mathcal{T}^{2}=1$, a new topological structure can be realized. This structure describes the case where a topological band-crossing traces out an extended loop in the Brillouin zone and is termed as nodal line (see Fig.~\ref{fig:FIG1}b). These nodal lines are topologically protected, with a non-zero $z_{2}$ number, and carry a Berry phase of $\pi$. This is in contrast to nodal lines in electronic band structures \cite{Young2015Dirac, Young2015Dirac, Fang2016, Bian2016Drumhead, Fu2019, Weyl_newfermions} where additional crystal symmetries are required to protect the band crossings as spin-orbit coupling (SOC) tends to lift the degeneracy away from high-symmetry locations. In this letter, we use first-principles calculations combined with meV-resolution inelastic x-ray scattering (IXS) to discover the first true $\mathcal{PT}$-symmetry protected nodal lines in the phonon spectrum of MoB$_{2}$. This material is a well known member of the super hard family of transition metal diborides \cite{tao2013enhanced}. This helical-shaped topological state is not restricted by any crystal symmetries and is prohibited in electronic structures. Interestingly, we find that the phononic nodal line induced surface bands are flat with high density-of-states (DOS) near the nodal energies, which might lead to electronic anomalies on the surface through electron-phonon coupling.

\begin{center}
\begin{figure}
\includegraphics[scale=1]{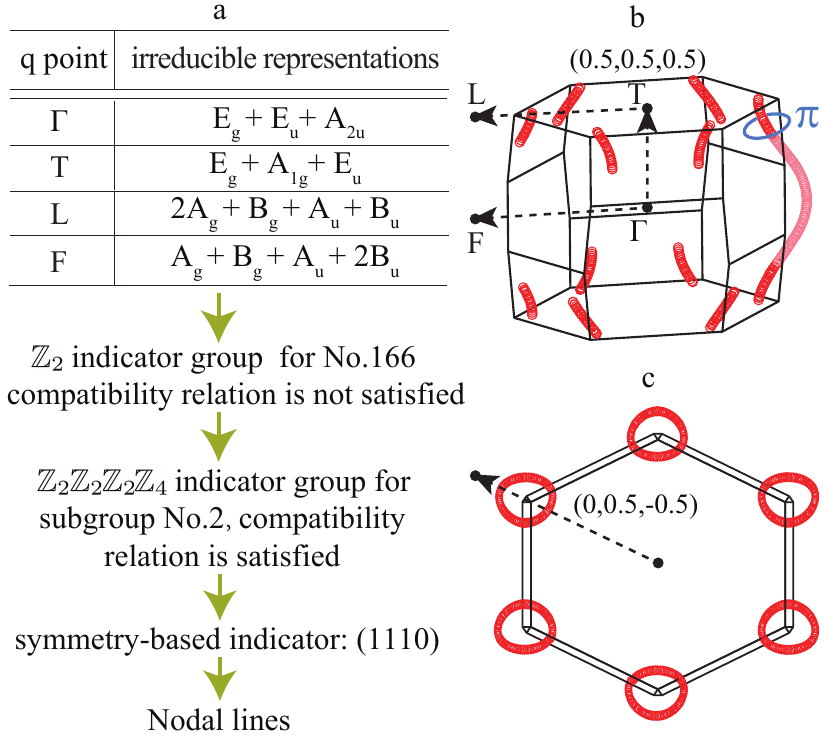}
\caption{(a) Diagnostic process for topological phonons by compatibility relation and symmetry-based indicator theories. (b) and (c) are the side view and top view (along the [111] direction) of the 30 meV nodal-line path in the BZ, respectively. The black dashed arrows in (b) indicate the IXS measurement trajectories. The projected trajectory in the (1,1,1)-plane is shown in (c). All nodal lines carry a quantized $\pi$ Berry phase. 
\label{fig:FIG2} }
\end{figure}
\end{center}

High quality single crystals of MoB$_{2}$ were grown via Al flux. The starting elements of Mo (99.5\%), B (99.9\%) and Al (99.99\%) were put into an alumina crucible, with a molar ratio of Mo : B : Al = 1 : 2.5 : 73.3. The mixture was heated up to 1773~K in a high-purity argon atmosphere and then slowly cooled down to 1173~K. The MoB$_{2}$ single crystals were separated from the Al flux using diluted hydrochloric acid. Phononic nodal lines were measured by IXS in a single crystal of MoB$_{2}$ of approximately 2$\times$3$\times$0.08~mm$^{3}$. The experiments were conducted at beamline 30-ID-C (HERIX \cite{Said2011,Toellner2011}) at the Advanced Photon Source (APS). The highly monochromatic x-ray beam of incident energy $E_{i}=23.7$~keV ($\lambda = 0.5226$~\AA{}) was focused on the sample with a beam cross section of $\sim35\times15$~\textmu{}$\mathrm{m}^{2}$ (horizontal $\times$ vertical). The total energy resolution of the monochromatic x-ray beam and analyzer crystals was $\Delta E\sim 1.5$ meV (full width at half maximum). The measurements were performed in reflection geometry. Typical counting times were in the range of 30 to 120 seconds per point in the energy scans at constant momentum transfer $\textbf{Q}$. The phonon dispersions of MoB$_{2}$ were calculated using the Vienna $ab$ $initio$ simulation package (VASP) \cite{CAL_VASP} that is based on density functional perturbation theory (DFPT) \cite{CAL_DFPT}. The exchange-correlation potential was treated within the generalized gradient approximation (GGA) of the Perdew-Burke-Ernzerhof variety \cite{DFT_GGA}. Prior to the phonon calculations, the crystal structure was relaxed until the residual force on each atom was less than 0.001~eV\AA$^{-1}$. The relaxed lattice parameters are in good agreement with experimental result \cite{Liu2014, ding2016crystal, supp}. To compute the surface states, we first calculate the second-order rank tensor of force constants in Cartesian coordinates based on DFPT, from which we obtain the tight-binding parameters for the bulk and surface atoms. We then construct the surface Green's function iteratively and take the imaginary part of the surface Green's function as the local density of states (LDOS) \cite{CAL_GREEN1,CAL_GREEN2,CAL_wanntools}.

MoB$_{2}$ has a centrosymmetric structure $R\bar{3}m$ (No.~166) \cite{Liu2014}, whose BZ, crystal structure and primitive cell are shown in Fig. \ref{fig:FIG1} (c)-(e), respectively. The Mo atomic layer sits between two different B atomic layers (labeled as B$_{1}$ and B$_{2}$ in Fig.~\ref{fig:FIG1}d, respectively) along the crystal $c$-direction or the [111] direction in the primitive-cell convention. The B$_{1}$ layer forms a planar quasi-2D honeycomb lattice while the B$_{2}$ layer is a buckled honeycomb network. Fig.~\ref{fig:FIG1}f shows the calculated phonon dispersion along the high-symmetry lines alongside the phonon DOS. The absence of negative phonon frequencies proves that the $R\bar{3}m$ structure of MoB$_{2}$ is stable, in agreement with an early study \cite{Liu2014}. As we show schematically in Fig.~\ref{fig:FIG1}b, the topological nodal line consists of two crossing bands. We thus inspect the phonon dispersions and find two sets of nodes: the lower energy nodes located near 7.25~THz (30 meV) and the higher energy nodes located around 23~THz (95 meV). As shown in Fig.~\ref{fig:FIG1}f, near these energies (indicated by dashed lines), the phonon DOS shows local minima, consistent with the nodal dispersion. As we go on to prove below, these nodes are protected by $\mathcal{PT}$ symmetry and form a helical topological nodal line through the BZ. 

\begin{figure*}
\includegraphics[scale=0.3]{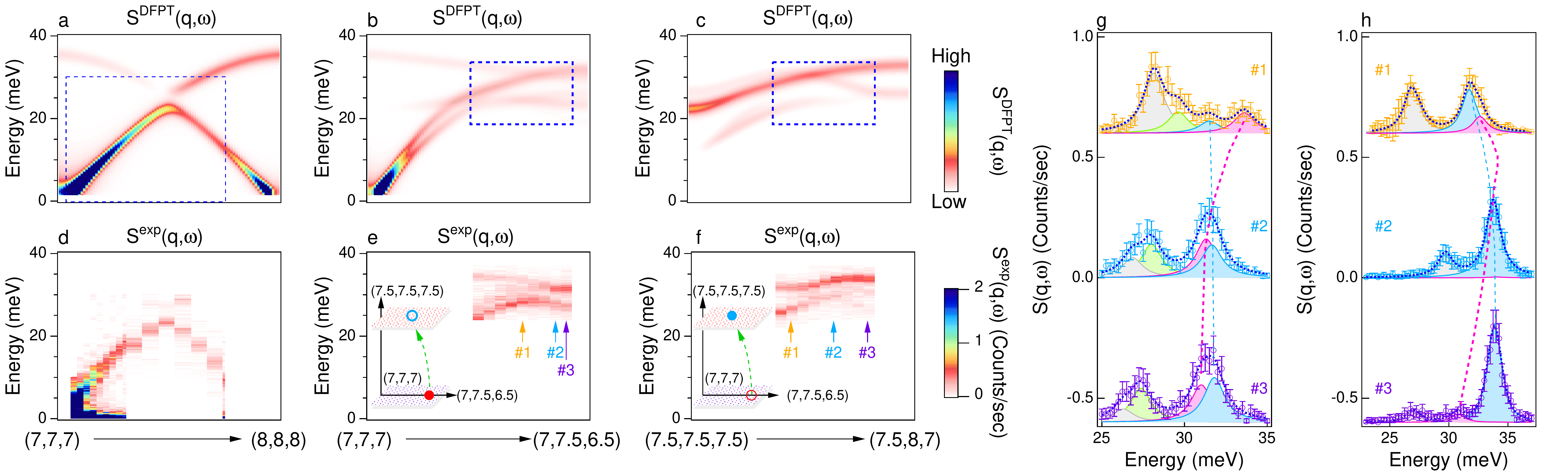}
\caption{(a-c) are the resolution convoluted phonon dynamic structural factor along different directions calculated $ab\ initio$. These momentum trajectories are equivalent to the black dashed trajectories shown in Fig.~\ref{fig:FIG2}b. The measured IXS data corresponding to the blue dashed squares in (a-c) are shown in (d-f), respectively. Insets in (e) and (f) schematically show the points intercepting between the helical nodal-line and high symmetry planes in momentum space. (g) and (h) show representative IXS spectra and fittings along the (7,7,7)$\rightarrow$(7,7.5,6.5) and (7.5,7.5,7.5)$\rightarrow$(7.5,8,7) directions, respectively. The scan numbers in (g) and (h) correspond to the positions shown in (e) and (f), respectively. Near the nodal lines (curve \#2 in g and h), the fitted peak separation is less than 0.3~meV, in agreement with the DFPT predicted band crossing.}
\label{fig:FIG3}
\end{figure*}

Symmetry-based indicator theories \cite{po2017symmetry,song2018quantitative,song2018diagnosis} and compatibility relations \cite{bradlyn2017TQC} are the main theoretical tools to diagnose topological states in electronic band structures. Here, we extend these methods to phonon spectra \cite{zhang2019catalogue,vergniory2019complete,tang2019comprehensive} and apply the diagnostic procedure to the 30 meV band crossing in MoB$_{2}$. As summarized in Fig.~\ref{fig:FIG2}a, the procedure starts with deducting the irreducible representations of the high-symmetry points $\Gamma$, $T$, $F$, and $L$ through first-principles calculations. We then check space group No.~166 with symmetry-based indicator theories and compatibility relations, which are available for all 230 space groups \cite{po2017symmetry,song2018quantitative,song2018diagnosis,bradlyn2017TQC}. We find that No.~166 has a nontrivial symmetry-based indicator group. For the 5 lowest energy phonon modes, however, the compatibility relation is not satisfied, suggesting band inversions between the 5th and 6th mode. To determine the topological structure, we use the indicator formulas of the subgroup No.~2, whose compatibility relation is satisfied for the first 5 lowest energy phonon modes. The symmetry-based indicator formulas of subgroup No.~2 can be written as:

\begin{equation}
z_{2,i}=\sum_{q\in \text{TRIM}\ \text{at}\ \{ q_{i}=\pi \}}\frac{N_{-}(q)-N_{+}(q)}{2}\ \text{mod}\ 2, i=1,2,3 
\label{indz2}
\end{equation}

\begin{equation}
z_{4}= \sum_{q\in \text{TRIM}}\frac{N_{-}(q)-N_{+}(q)}{2}\ \text{mod}\ 4
\label{indz4}
\end{equation}
where $\text{TRIM}$ stands for time-reversal invariant momenta. $N_{+}(q)$ and $N_{-}(q)$ represent the number of bands that are even or odd under $\mathcal{P}$ symmetry in $q$, respectively. Eq.~\ref{indz2} and Eq.~\ref{indz4} give the types, position, number, and topological charge of the band crossings in MoB$_{2}$ \cite{song2018diagnosis}. Finally we use the indicator of $z_{2,1}z_{2,2}z_{2,3}z_{2,4}z_{4}=(1110)$ for subgroup No.~2, which corresponds to 2 $\text{mod}$ 4 nodal lines along the [111] direction in the BZ. These nodal lines carry a $\pi$ Berry phase and are robust against small perturbations that do not break $\mathcal{PT}$ symmetry. 

We perform meV-resolution IXS measurement to directly probe the bulk phonon band crossings of MoB$_2$ and the presence of the nodal line. IXS measures the phonon dynamic structure factor, $S(\mathbi{Q},\omega)$, which can be formulated as

\begin{equation}
S(\Q,\omega)=\sum_{j}|F(\tau,\mathbi{q},j)|^{2}/(1-e^{-\hbar\omega_{\mathbi{q}j}/k_{B}T})\delta(\omega-\omega_{\mathbi{q}j})
\label{S_qw}
\end{equation}
where $\mathbi{Q}=\mathbi{q}+\tau$, $\mathbi{q}$, and $\tau$ are the total momentum transfer, reduced momentum transfer in the first BZ and the reciprocal lattice vector, respectively. $\omega_{\mathbi{q}j}$ is the $j$th energy of phonon mode at momentum $\mathbi{q}$ \cite{Baron2015,Miao2018}. $|F(\tau,\mathbi{q},j)|^{2}$ is given by

\begin{equation}
|F(\tau,\mathbi{q},j)|^{2}=\frac{1}{\omega_{\mathbi{q}j}}\left| \sum_{d}\frac{f_{d}(\Q)}{\sqrt{2M_{d}}}\Q\cdot\textbf{e}_{\mathbi{q}jd}e^{i\Q\cdot\mathbi{r}_{d}}\right|^{2}
\label{F_qj}
\end{equation}
where $f_{d}$ and $M_{d}$ are the atomic form-factor and atomic mass, respectively. The index $d$ denotes atoms inside the primitive cell. Vector $\textbf{e}_{\mathbi{q}jd}$ is the eigenvector of the phonon. As shown in Fig.~\ref{fig:FIG1}f, the nodal lines intercept the high-symmetry directions: ($h$, $h$, $h$)$\rightarrow$($h$, $h$+0.5, $h$-0.5) and ($h$+0.5, $h$+0.5, $h$+0.5)$\rightarrow$($h$+0.5,$h$+1,$h$), where $h$ is an integer. Since $S(\Q,\omega)$ varies strongly in different BZs, we first calculate the resolution convoluted $S^{\text{DFPT}}(\Q,\omega)$ and decide to perform our measurement along the (7,7,7)$\rightarrow$(8,8,8), (7,7,7)$\rightarrow$(7,7.5, 6.5) and (7.5,7.5,7.5)$\rightarrow$(7.5,8,7) directions (Fig.~\ref{fig:FIG3} (a-c)). The IXS data corresponding to the blue dashed squares in Fig.~\ref{fig:FIG3} (a-c) are shown in Fig.~\ref{fig:FIG3} (d-f), respectively. To precisely determine the nodal-line loci along the (7,7,7)$\rightarrow$(7,7.5,6.5) and (7.5,7.5,7.5)$\rightarrow$(7.5,8,7) directions, we extract the phonon peak positions by fitting the IXS spectra \cite{supp}. Figure~\ref{fig:FIG3}g and h show representative fittings. The number of phonon peaks that are used in the fitting energy window are based on the theoretical calculations shown in Fig.~\ref{fig:FIG3}b and c. The high energy-resolution of our measurement enable us to resolve band splitting larger than 0.5~meV. Within this limit, we find that along the (7,7,7)$\rightarrow$(7,7.5,6.5) direction, the topological band-crossing is observed near scan \#2, corresponding to $Q=(7,7,7)+0.84\times(0,0.5,-0.5)$. Along the (7.5,7.5,7.5)$\rightarrow$(7.5,8,7) direction, the topological band-crossing moves closer to the zone center and locates near $Q=(7.5, 7.5, 7.5)+0.6\times(0,0.5,-0.5)$, consistent with the first principles calculations. In this way, we directly trace out the extended nature of the band crossing that describes the nodal line, demonstrating the realization of this state in MoB$_{2}$.

Having experimentally established the topologically non-trivial band crossings along the high-symmetry lines, we perform complete phonon calculations in the entire BZ, and map out the two helical nodal lines winding along the [111] direction. The validity of our calculations is supported by the excellent agreement with the measurements shown in Fig.~\ref{fig:FIG3}. Since these topological states are not restricted by any additional crystal symmetries, the nodal-line path does not overlap with the high-symmetry directions in momentum space except when it intercepts with the (0,0,0) and (0.5,0.5,0.5) planes, as shown in Fig.~\ref{fig:FIG2} (b) and (c). This unique helical shape cannot be realized in electronic structures, as the presence of finite SOC will lift the degeneracy except at high-symmetry locations. Using the same diagnostic procedure, we find that the band crossing near 95 meV shares a similar topology with those near 30 meV. Its topological invariant is found to be $z_{2,1}z_{2,2}z_{2,3}z_{2,4}z_{4}=(1112)$ \cite{supp}. By examining the eigenvectors of the crossing bands, we find that the 30 meV nodal lines are dominated by Mo-vibrations on the triangular lattice, while the 95 meV nodal-lines are mainly composed of B$_1$ and B$_2$ vibrations on the planar and buckled hexagonal lattice, respectively. The same symmetry of the triangular lattice and the buckled hexagonal lattice explains the similar topological nature of the low and high energy nodal lines. 

\begin{figure}
\includegraphics[scale=1.1]{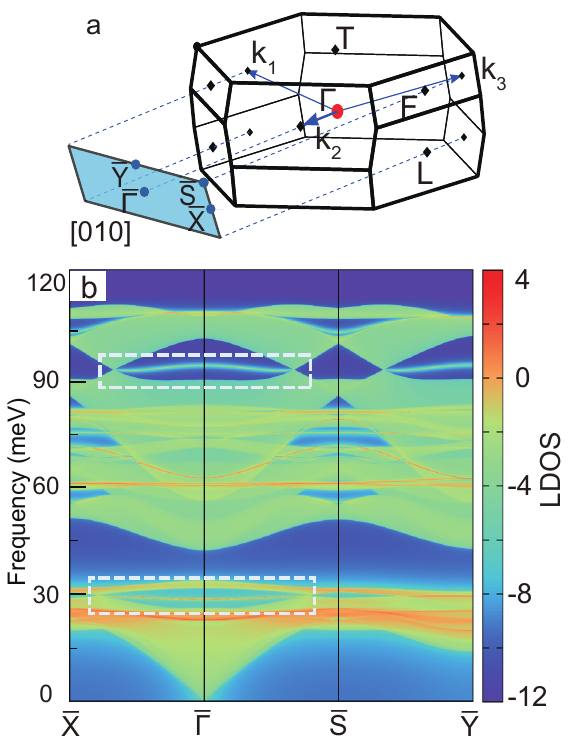}
\caption{(a) The bulk BZ and the surface BZ along the [010] direction of MoB$_{2}$. (b) shows the calculated surface phonon modes together with the projected bulk phonon dispersions along the [010] direction. The colorscale represents the value of the imaginary part of LDOS on a log scale. The drumhead surface phonon states (connecting the node-line crossing points in side the white dashed rectangles) have the highest LDOS as manifested by the sharp orange lines on top of the projected green/yellow bulk bands.
\label{fig:FIG4}}
\end{figure}

The combined theoretical analysis and bulk-sensitive IXS measurements demonstrate the first realization of $\mathcal{PT}$-symmetry protected helical nodal-lines in MoB$_{2}$. Similar to other topological structures, these new bulk topological states induce phononic boundary modes on the surface that are robust against local disorders. Figure~\ref{fig:FIG4} shows the LDOS on a log scale. The intensity is calculated by the Green's function method with a semi-finite system projected on the $(010)$ surface. Interestingly, unlike the Weyl points, which induce dispersive helicoid edge modes \cite{miao2018observation,zhang2018double}, the surface states arising from the topological nodal lines form a drumhead shape. In contrast to the bulk phonon DOS minima near 30 meV and 95 meV, the flat surface modes yield high surface DOS near these energies. Such high surface phonon DOS may thus induce surface electronic structure anomalies through electron-phonon coupling \cite{Valla1999, Tamtogl2018}. It would be interesting for future studies to explore such electronic anomalies via angle-resolved photoemission spectroscopy, which may provide an alternative route to justify the presence of nodal-line edge modes.

In summary, using first principles calculations and meV-resolution IXS, we show the first realization of $\mathcal{PT}$ symmetry protected helical nodal lines in the phonon spectrum of MoB$_{2}$. Our work provides a blueprint for systematic identification and verification of topological Bosonic states in crystalline materials.

\begin{acknowledgements}
This material is based upon work supported by the U.S. Department of Energy, Office of Basic Energy Sciences, Early Career Award Program under Award No. 1047478. Work at Brookhaven National Laboratory was supported by the U.S. Department of Energy, Office of Science, Office of Basic Energy Sciences, under Contract No. DE-SC0012704. The IXS experiment were performed at 30ID in the Advanced Photon Source, a U.S. Department of Energy (DOE) Office of Science User Facility operated for the DOE Office of Science by Argonne National Laboratory under Contract No. DE-AC02-06CH11357. Argonne National Laboratory’s contribution is based upon work supported by Laboratory Directed Research and Development (LDRD) funding from Argonne National Laboratory, provided by the Director, Office of Science, of the U.S. Department of Energy under Contract No. DE-AC02-06CH11357.We acknowledge the supports from the National Key Research and Development Program of China (Grant No. 2016YFA0300600), the National Natural Science Foundation of China (Grant No. 11674369). Q.W. and H.C.L. Acknowledge the support from the National Key R\&D Program of China (Grants No. 2016YFA0300504), the National Natural Science Foundation of China (No. 11574394, 11774423,11822412), the Fundamental Research Funds for the Central Universities, and the Research Funds of Renmin University of China (RUC) (15XNLQ07, 18XNLG14, 19XNLG17). T. T. Zhang and H. Miao are contributed equally to this work.

\end{acknowledgements}

\bibliography{reference.bib}

\begin{thebibliography}{37}%
\makeatletter
\providecommand \@ifxundefined [1]{%
 \@ifx{#1\undefined}
}%
\providecommand \@ifnum [1]{%
 \ifnum #1\expandafter \@firstoftwo
 \else \expandafter \@secondoftwo
 \fi
}%
\providecommand \@ifx [1]{%
 \ifx #1\expandafter \@firstoftwo
 \else \expandafter \@secondoftwo
 \fi
}%
\providecommand \natexlab [1]{#1}%
\providecommand \enquote  [1]{``#1''}%
\providecommand \bibnamefont  [1]{#1}%
\providecommand \bibfnamefont [1]{#1}%
\providecommand \citenamefont [1]{#1}%
\providecommand \href@noop [0]{\@secondoftwo}%
\providecommand \href [0]{\begingroup \@sanitize@url \@href}%
\providecommand \@href[1]{\@@startlink{#1}\@@href}%
\providecommand \@@href[1]{\endgroup#1\@@endlink}%
\providecommand \@sanitize@url [0]{\catcode `\\12\catcode `\$12\catcode
  `\&12\catcode `\#12\catcode `\^12\catcode `\_12\catcode `\%12\relax}%
\providecommand \@@startlink[1]{}%
\providecommand \@@endlink[0]{}%
\providecommand \url  [0]{\begingroup\@sanitize@url \@url }%
\providecommand \@url [1]{\endgroup\@href {#1}{\urlprefix }}%
\providecommand \urlprefix  [0]{URL }%
\providecommand \Eprint [0]{\href }%
\providecommand \doibase [0]{http://dx.doi.org/}%
\providecommand \selectlanguage [0]{\@gobble}%
\providecommand \bibinfo  [0]{\@secondoftwo}%
\providecommand \bibfield  [0]{\@secondoftwo}%
\providecommand \translation [1]{[#1]}%
\providecommand \BibitemOpen [0]{}%
\providecommand \bibitemStop [0]{}%
\providecommand \bibitemNoStop [0]{.\EOS\space}%
\providecommand \EOS [0]{\spacefactor3000\relax}%
\providecommand \BibitemShut  [1]{\csname bibitem#1\endcsname}%
\let\auto@bib@innerbib\@empty
\bibitem [{\citenamefont {Po}\ \emph {et~al.}(2017)\citenamefont {Po},
  \citenamefont {Vishwanath},\ and\ \citenamefont {Watanabe}}]{po2017symmetry}%
  \BibitemOpen
  \bibfield  {author} {\bibinfo {author} {\bibfnamefont {H.~C.}\ \bibnamefont
  {Po}}, \bibinfo {author} {\bibfnamefont {A.}~\bibnamefont {Vishwanath}}, \
  and\ \bibinfo {author} {\bibfnamefont {H.}~\bibnamefont {Watanabe}},\
  }\href@noop {} {\bibfield  {journal} {\bibinfo  {journal} {Nature
  Communications}\ }\textbf {\bibinfo {volume} {8}},\ \bibinfo {pages} {50}
  (\bibinfo {year} {2017})}\BibitemShut {NoStop}%
\bibitem [{\citenamefont {Song}\ \emph
  {et~al.}(2018{\natexlab{a}})\citenamefont {Song}, \citenamefont {Zhang},
  \citenamefont {Fang},\ and\ \citenamefont {Fang}}]{song2018quantitative}%
  \BibitemOpen
  \bibfield  {author} {\bibinfo {author} {\bibfnamefont {Z.}~\bibnamefont
  {Song}}, \bibinfo {author} {\bibfnamefont {T.}~\bibnamefont {Zhang}},
  \bibinfo {author} {\bibfnamefont {Z.}~\bibnamefont {Fang}}, \ and\ \bibinfo
  {author} {\bibfnamefont {C.}~\bibnamefont {Fang}},\ }\href@noop {} {\bibfield
   {journal} {\bibinfo  {journal} {Nature communications}\ }\textbf {\bibinfo
  {volume} {9}},\ \bibinfo {pages} {3530} (\bibinfo {year}
  {2018}{\natexlab{a}})}\BibitemShut {NoStop}%
\bibitem [{\citenamefont {Song}\ \emph
  {et~al.}(2018{\natexlab{b}})\citenamefont {Song}, \citenamefont {Zhang},\
  and\ \citenamefont {Fang}}]{song2018diagnosis}%
  \BibitemOpen
  \bibfield  {author} {\bibinfo {author} {\bibfnamefont {Z.}~\bibnamefont
  {Song}}, \bibinfo {author} {\bibfnamefont {T.}~\bibnamefont {Zhang}}, \ and\
  \bibinfo {author} {\bibfnamefont {C.}~\bibnamefont {Fang}},\ }\href@noop {}
  {\bibfield  {journal} {\bibinfo  {journal} {Physical Review X}\ }\textbf
  {\bibinfo {volume} {8}},\ \bibinfo {pages} {031069} (\bibinfo {year}
  {2018}{\natexlab{b}})}\BibitemShut {NoStop}%
\bibitem [{\citenamefont {Bradlyn}\ \emph {et~al.}(2017)\citenamefont
  {Bradlyn}, \citenamefont {Elcoro}, \citenamefont {Cano}, \citenamefont
  {Vergniory}, \citenamefont {Wang}, \citenamefont {Felser}, \citenamefont
  {Aroyo},\ and\ \citenamefont {Bernevig}}]{bradlyn2017TQC}%
  \BibitemOpen
  \bibfield  {author} {\bibinfo {author} {\bibfnamefont {B.}~\bibnamefont
  {Bradlyn}}, \bibinfo {author} {\bibfnamefont {L.}~\bibnamefont {Elcoro}},
  \bibinfo {author} {\bibfnamefont {J.}~\bibnamefont {Cano}}, \bibinfo {author}
  {\bibfnamefont {M.}~\bibnamefont {Vergniory}}, \bibinfo {author}
  {\bibfnamefont {Z.}~\bibnamefont {Wang}}, \bibinfo {author} {\bibfnamefont
  {C.}~\bibnamefont {Felser}}, \bibinfo {author} {\bibfnamefont
  {M.}~\bibnamefont {Aroyo}}, \ and\ \bibinfo {author} {\bibfnamefont {B.~A.}\
  \bibnamefont {Bernevig}},\ }\href@noop {} {\bibfield  {journal} {\bibinfo
  {journal} {Nature}\ }\textbf {\bibinfo {volume} {547}},\ \bibinfo {pages}
  {298} (\bibinfo {year} {2017})}\BibitemShut {NoStop}%
\bibitem [{\citenamefont {Wang}\ \emph {et~al.}(2015)\citenamefont {Wang},
  \citenamefont {Lu},\ and\ \citenamefont {Bertoldi}}]{Wang2015}%
  \BibitemOpen
  \bibfield  {author} {\bibinfo {author} {\bibfnamefont {P.}~\bibnamefont
  {Wang}}, \bibinfo {author} {\bibfnamefont {L.}~\bibnamefont {Lu}}, \ and\
  \bibinfo {author} {\bibfnamefont {K.}~\bibnamefont {Bertoldi}},\ }\href
  {\doibase 10.1103/PhysRevLett.115.104302} {\bibfield  {journal} {\bibinfo
  {journal} {Phys. Rev. Lett.}\ }\textbf {\bibinfo {volume} {115}},\ \bibinfo
  {pages} {104302} (\bibinfo {year} {2015})}\BibitemShut {NoStop}%
\bibitem [{\citenamefont {Stenull}\ \emph {et~al.}(2016)\citenamefont
  {Stenull}, \citenamefont {Kane},\ and\ \citenamefont
  {Lubensky}}]{Stenull2016}%
  \BibitemOpen
  \bibfield  {author} {\bibinfo {author} {\bibfnamefont {O.}~\bibnamefont
  {Stenull}}, \bibinfo {author} {\bibfnamefont {C.~L.}\ \bibnamefont {Kane}}, \
  and\ \bibinfo {author} {\bibfnamefont {T.~C.}\ \bibnamefont {Lubensky}},\
  }\href {\doibase 10.1103/PhysRevLett.117.068001} {\bibfield  {journal}
  {\bibinfo  {journal} {Phys. Rev. Lett.}\ }\textbf {\bibinfo {volume} {117}},\
  \bibinfo {pages} {068001} (\bibinfo {year} {2016})}\BibitemShut {NoStop}%
\bibitem [{\citenamefont {Chiu}\ \emph {et~al.}(2016)\citenamefont {Chiu},
  \citenamefont {Teo}, \citenamefont {Schnyder},\ and\ \citenamefont
  {Ryu}}]{Top_classifacation_symmetry}%
  \BibitemOpen
  \bibfield  {author} {\bibinfo {author} {\bibfnamefont {C.-K.}\ \bibnamefont
  {Chiu}}, \bibinfo {author} {\bibfnamefont {J.~C.~Y.}\ \bibnamefont {Teo}},
  \bibinfo {author} {\bibfnamefont {A.~P.}\ \bibnamefont {Schnyder}}, \ and\
  \bibinfo {author} {\bibfnamefont {S.}~\bibnamefont {Ryu}},\ }\href {\doibase
  10.1103/RevModPhys.88.035005} {\bibfield  {journal} {\bibinfo  {journal}
  {Rev. Mod. Phys.}\ }\textbf {\bibinfo {volume} {88}},\ \bibinfo {pages}
  {035005} (\bibinfo {year} {2016})}\BibitemShut {NoStop}%
\bibitem [{\citenamefont {Miao}\ \emph
  {et~al.}(2018{\natexlab{a}})\citenamefont {Miao}, \citenamefont {Zhang},
  \citenamefont {Wang}, \citenamefont {Meyers}, \citenamefont {Said},
  \citenamefont {Wang}, \citenamefont {Shi}, \citenamefont {Weng},
  \citenamefont {Fang},\ and\ \citenamefont {Dean}}]{miao2018observation}%
  \BibitemOpen
  \bibfield  {author} {\bibinfo {author} {\bibfnamefont {H.}~\bibnamefont
  {Miao}}, \bibinfo {author} {\bibfnamefont {T.}~\bibnamefont {Zhang}},
  \bibinfo {author} {\bibfnamefont {L.}~\bibnamefont {Wang}}, \bibinfo {author}
  {\bibfnamefont {D.}~\bibnamefont {Meyers}}, \bibinfo {author} {\bibfnamefont
  {A.}~\bibnamefont {Said}}, \bibinfo {author} {\bibfnamefont {Y.}~\bibnamefont
  {Wang}}, \bibinfo {author} {\bibfnamefont {Y.}~\bibnamefont {Shi}}, \bibinfo
  {author} {\bibfnamefont {H.}~\bibnamefont {Weng}}, \bibinfo {author}
  {\bibfnamefont {Z.}~\bibnamefont {Fang}}, \ and\ \bibinfo {author}
  {\bibfnamefont {M.}~\bibnamefont {Dean}},\ }\href@noop {} {\bibfield
  {journal} {\bibinfo  {journal} {Physical review letters}\ }\textbf {\bibinfo
  {volume} {121}},\ \bibinfo {pages} {035302} (\bibinfo {year}
  {2018}{\natexlab{a}})}\BibitemShut {NoStop}%
\bibitem [{\citenamefont {Zhang}\ \emph {et~al.}(2018)\citenamefont {Zhang},
  \citenamefont {Song}, \citenamefont {Alexandradinata}, \citenamefont {Weng},
  \citenamefont {Fang}, \citenamefont {Lu},\ and\ \citenamefont
  {Fang}}]{zhang2018double}%
  \BibitemOpen
  \bibfield  {author} {\bibinfo {author} {\bibfnamefont {T.}~\bibnamefont
  {Zhang}}, \bibinfo {author} {\bibfnamefont {Z.}~\bibnamefont {Song}},
  \bibinfo {author} {\bibfnamefont {A.}~\bibnamefont {Alexandradinata}},
  \bibinfo {author} {\bibfnamefont {H.}~\bibnamefont {Weng}}, \bibinfo {author}
  {\bibfnamefont {C.}~\bibnamefont {Fang}}, \bibinfo {author} {\bibfnamefont
  {L.}~\bibnamefont {Lu}}, \ and\ \bibinfo {author} {\bibfnamefont
  {Z.}~\bibnamefont {Fang}},\ }\href@noop {} {\bibfield  {journal} {\bibinfo
  {journal} {Physical review letters}\ }\textbf {\bibinfo {volume} {120}},\
  \bibinfo {pages} {016401} (\bibinfo {year} {2018})}\BibitemShut {NoStop}%
\bibitem [{\citenamefont {Li}\ \emph {et~al.}(2017)\citenamefont {Li},
  \citenamefont {Li}, \citenamefont {Hu}, \citenamefont {Li},\ and\
  \citenamefont {Fang}}]{li2017dirac}%
  \BibitemOpen
  \bibfield  {author} {\bibinfo {author} {\bibfnamefont {K.}~\bibnamefont
  {Li}}, \bibinfo {author} {\bibfnamefont {C.}~\bibnamefont {Li}}, \bibinfo
  {author} {\bibfnamefont {J.}~\bibnamefont {Hu}}, \bibinfo {author}
  {\bibfnamefont {Y.}~\bibnamefont {Li}}, \ and\ \bibinfo {author}
  {\bibfnamefont {C.}~\bibnamefont {Fang}},\ }\href@noop {} {\bibfield
  {journal} {\bibinfo  {journal} {Physical review letters}\ }\textbf {\bibinfo
  {volume} {119}},\ \bibinfo {pages} {247202} (\bibinfo {year}
  {2017})}\BibitemShut {NoStop}%
\bibitem [{\citenamefont {Yao}\ \emph {et~al.}(2018)\citenamefont {Yao},
  \citenamefont {Li}, \citenamefont {Wang}, \citenamefont {Xue}, \citenamefont
  {Dan}, \citenamefont {Iida}, \citenamefont {Kamazawa}, \citenamefont {Li},
  \citenamefont {Fang},\ and\ \citenamefont {Li}}]{yao2018topological}%
  \BibitemOpen
  \bibfield  {author} {\bibinfo {author} {\bibfnamefont {W.}~\bibnamefont
  {Yao}}, \bibinfo {author} {\bibfnamefont {C.}~\bibnamefont {Li}}, \bibinfo
  {author} {\bibfnamefont {L.}~\bibnamefont {Wang}}, \bibinfo {author}
  {\bibfnamefont {S.}~\bibnamefont {Xue}}, \bibinfo {author} {\bibfnamefont
  {Y.}~\bibnamefont {Dan}}, \bibinfo {author} {\bibfnamefont {K.}~\bibnamefont
  {Iida}}, \bibinfo {author} {\bibfnamefont {K.}~\bibnamefont {Kamazawa}},
  \bibinfo {author} {\bibfnamefont {K.}~\bibnamefont {Li}}, \bibinfo {author}
  {\bibfnamefont {C.}~\bibnamefont {Fang}}, \ and\ \bibinfo {author}
  {\bibfnamefont {Y.}~\bibnamefont {Li}},\ }\href@noop {} {\bibfield  {journal}
  {\bibinfo  {journal} {Nature Physics}\ }\textbf {\bibinfo {volume} {14}},\
  \bibinfo {pages} {1011} (\bibinfo {year} {2018})}\BibitemShut {NoStop}%
\bibitem [{\citenamefont {Fang}\ \emph {et~al.}(2003)\citenamefont {Fang},
  \citenamefont {Nagaosa}, \citenamefont {Takahashi}, \citenamefont {Asamitsu},
  \citenamefont {Mathieu}, \citenamefont {Ogasawara}, \citenamefont {Yamada},
  \citenamefont {Kawasaki}, \citenamefont {Tokura},\ and\ \citenamefont
  {Terakura}}]{fang2003anomalous}%
  \BibitemOpen
  \bibfield  {author} {\bibinfo {author} {\bibfnamefont {Z.}~\bibnamefont
  {Fang}}, \bibinfo {author} {\bibfnamefont {N.}~\bibnamefont {Nagaosa}},
  \bibinfo {author} {\bibfnamefont {K.~S.}\ \bibnamefont {Takahashi}}, \bibinfo
  {author} {\bibfnamefont {A.}~\bibnamefont {Asamitsu}}, \bibinfo {author}
  {\bibfnamefont {R.}~\bibnamefont {Mathieu}}, \bibinfo {author} {\bibfnamefont
  {T.}~\bibnamefont {Ogasawara}}, \bibinfo {author} {\bibfnamefont
  {H.}~\bibnamefont {Yamada}}, \bibinfo {author} {\bibfnamefont
  {M.}~\bibnamefont {Kawasaki}}, \bibinfo {author} {\bibfnamefont
  {Y.}~\bibnamefont {Tokura}}, \ and\ \bibinfo {author} {\bibfnamefont
  {K.}~\bibnamefont {Terakura}},\ }\href@noop {} {\bibfield  {journal}
  {\bibinfo  {journal} {Science}\ }\textbf {\bibinfo {volume} {302}},\ \bibinfo
  {pages} {92} (\bibinfo {year} {2003})}\BibitemShut {NoStop}%
\bibitem [{\citenamefont {Armitage}\ \emph {et~al.}(2018)\citenamefont
  {Armitage}, \citenamefont {Mele},\ and\ \citenamefont
  {Vishwanath}}]{armitage2018weyl}%
  \BibitemOpen
  \bibfield  {author} {\bibinfo {author} {\bibfnamefont {N.}~\bibnamefont
  {Armitage}}, \bibinfo {author} {\bibfnamefont {E.}~\bibnamefont {Mele}}, \
  and\ \bibinfo {author} {\bibfnamefont {A.}~\bibnamefont {Vishwanath}},\
  }\href@noop {} {\bibfield  {journal} {\bibinfo  {journal} {Reviews of Modern
  Physics}\ }\textbf {\bibinfo {volume} {90}},\ \bibinfo {pages} {015001}
  (\bibinfo {year} {2018})}\BibitemShut {NoStop}%
\bibitem [{\citenamefont {Young}\ and\ \citenamefont
  {Kane}(2015)}]{Young2015Dirac}%
  \BibitemOpen
  \bibfield  {author} {\bibinfo {author} {\bibfnamefont {S.~M.}\ \bibnamefont
  {Young}}\ and\ \bibinfo {author} {\bibfnamefont {C.~L.}\ \bibnamefont
  {Kane}},\ }\href {\doibase 10.1103/PhysRevLett.115.126803} {\bibfield
  {journal} {\bibinfo  {journal} {Phys. Rev. Lett.}\ }\textbf {\bibinfo
  {volume} {115}},\ \bibinfo {pages} {126803} (\bibinfo {year}
  {2015})}\BibitemShut {NoStop}%
\bibitem [{\citenamefont {Fang}\ \emph {et~al.}(2016)\citenamefont {Fang},
  \citenamefont {Weng}, \citenamefont {Dai},\ and\ \citenamefont
  {Fang}}]{Fang2016}%
  \BibitemOpen
  \bibfield  {author} {\bibinfo {author} {\bibfnamefont {C.}~\bibnamefont
  {Fang}}, \bibinfo {author} {\bibfnamefont {H.}~\bibnamefont {Weng}}, \bibinfo
  {author} {\bibfnamefont {X.}~\bibnamefont {Dai}}, \ and\ \bibinfo {author}
  {\bibfnamefont {Z.}~\bibnamefont {Fang}},\ }\href {\doibase
  10.1088/1674-1056/25/11/117106} {\bibfield  {journal} {\bibinfo  {journal}
  {Chinese Physics B}\ }\textbf {\bibinfo {volume} {25}},\ \bibinfo {pages}
  {117106} (\bibinfo {year} {2016})}\BibitemShut {NoStop}%
\bibitem [{\citenamefont {Bian}\ \emph {et~al.}(2016)\citenamefont {Bian},
  \citenamefont {Chang}, \citenamefont {Zheng}, \citenamefont {Velury},
  \citenamefont {Xu}, \citenamefont {Neupert}, \citenamefont {Chiu},
  \citenamefont {Huang}, \citenamefont {Sanchez}, \citenamefont {Belopolski},
  \citenamefont {Alidoust}, \citenamefont {Chen}, \citenamefont {Chang},
  \citenamefont {Bansil}, \citenamefont {Jeng}, \citenamefont {Lin},\ and\
  \citenamefont {Hasan}}]{Bian2016Drumhead}%
  \BibitemOpen
  \bibfield  {author} {\bibinfo {author} {\bibfnamefont {G.}~\bibnamefont
  {Bian}}, \bibinfo {author} {\bibfnamefont {T.-R.}\ \bibnamefont {Chang}},
  \bibinfo {author} {\bibfnamefont {H.}~\bibnamefont {Zheng}}, \bibinfo
  {author} {\bibfnamefont {S.}~\bibnamefont {Velury}}, \bibinfo {author}
  {\bibfnamefont {S.-Y.}\ \bibnamefont {Xu}}, \bibinfo {author} {\bibfnamefont
  {T.}~\bibnamefont {Neupert}}, \bibinfo {author} {\bibfnamefont {C.-K.}\
  \bibnamefont {Chiu}}, \bibinfo {author} {\bibfnamefont {S.-M.}\ \bibnamefont
  {Huang}}, \bibinfo {author} {\bibfnamefont {D.~S.}\ \bibnamefont {Sanchez}},
  \bibinfo {author} {\bibfnamefont {I.}~\bibnamefont {Belopolski}}, \bibinfo
  {author} {\bibfnamefont {N.}~\bibnamefont {Alidoust}}, \bibinfo {author}
  {\bibfnamefont {P.-J.}\ \bibnamefont {Chen}}, \bibinfo {author}
  {\bibfnamefont {G.}~\bibnamefont {Chang}}, \bibinfo {author} {\bibfnamefont
  {A.}~\bibnamefont {Bansil}}, \bibinfo {author} {\bibfnamefont {H.-T.}\
  \bibnamefont {Jeng}}, \bibinfo {author} {\bibfnamefont {H.}~\bibnamefont
  {Lin}}, \ and\ \bibinfo {author} {\bibfnamefont {M.~Z.}\ \bibnamefont
  {Hasan}},\ }\href {\doibase 10.1103/PhysRevB.93.121113} {\bibfield  {journal}
  {\bibinfo  {journal} {Phys. Rev. B}\ }\textbf {\bibinfo {volume} {93}},\
  \bibinfo {pages} {121113} (\bibinfo {year} {2016})}\BibitemShut {NoStop}%
\bibitem [{\citenamefont {Fu}\ \emph {et~al.}(2019)\citenamefont {Fu},
  \citenamefont {Yi}, \citenamefont {Zhang}, \citenamefont {Caputo},
  \citenamefont {Ma}, \citenamefont {Gao}, \citenamefont {Lv}, \citenamefont
  {Kong}, \citenamefont {Huang}, \citenamefont {Richard}, \citenamefont {Shi},
  \citenamefont {Strocov}, \citenamefont {Fang}, \citenamefont {Weng},
  \citenamefont {Shi}, \citenamefont {Qian},\ and\ \citenamefont
  {Ding}}]{Fu2019}%
  \BibitemOpen
  \bibfield  {author} {\bibinfo {author} {\bibfnamefont {B.-B.}\ \bibnamefont
  {Fu}}, \bibinfo {author} {\bibfnamefont {C.-J.}\ \bibnamefont {Yi}}, \bibinfo
  {author} {\bibfnamefont {T.-T.}\ \bibnamefont {Zhang}}, \bibinfo {author}
  {\bibfnamefont {M.}~\bibnamefont {Caputo}}, \bibinfo {author} {\bibfnamefont
  {J.-Z.}\ \bibnamefont {Ma}}, \bibinfo {author} {\bibfnamefont
  {X.}~\bibnamefont {Gao}}, \bibinfo {author} {\bibfnamefont {B.~Q.}\
  \bibnamefont {Lv}}, \bibinfo {author} {\bibfnamefont {L.-Y.}\ \bibnamefont
  {Kong}}, \bibinfo {author} {\bibfnamefont {Y.-B.}\ \bibnamefont {Huang}},
  \bibinfo {author} {\bibfnamefont {P.}~\bibnamefont {Richard}}, \bibinfo
  {author} {\bibfnamefont {M.}~\bibnamefont {Shi}}, \bibinfo {author}
  {\bibfnamefont {V.~N.}\ \bibnamefont {Strocov}}, \bibinfo {author}
  {\bibfnamefont {C.}~\bibnamefont {Fang}}, \bibinfo {author} {\bibfnamefont
  {H.-M.}\ \bibnamefont {Weng}}, \bibinfo {author} {\bibfnamefont {Y.-G.}\
  \bibnamefont {Shi}}, \bibinfo {author} {\bibfnamefont {T.}~\bibnamefont
  {Qian}}, \ and\ \bibinfo {author} {\bibfnamefont {H.}~\bibnamefont {Ding}},\
  }\href {\doibase 10.1126/sciadv.aau6459} {\bibfield  {journal} {\bibinfo
  {journal} {Science Advances}\ }\textbf {\bibinfo {volume} {5}} (\bibinfo
  {year} {2019}),\ 10.1126/sciadv.aau6459}\BibitemShut {NoStop}%
\bibitem [{\citenamefont {Bradlyn}\ \emph {et~al.}(2016)\citenamefont
  {Bradlyn}, \citenamefont {Cano}, \citenamefont {Wang}, \citenamefont
  {Vergniory}, \citenamefont {Felser}, \citenamefont {Cava},\ and\
  \citenamefont {Bernevig}}]{Weyl_newfermions}%
  \BibitemOpen
  \bibfield  {author} {\bibinfo {author} {\bibfnamefont {B.}~\bibnamefont
  {Bradlyn}}, \bibinfo {author} {\bibfnamefont {J.}~\bibnamefont {Cano}},
  \bibinfo {author} {\bibfnamefont {Z.}~\bibnamefont {Wang}}, \bibinfo {author}
  {\bibfnamefont {M.}~\bibnamefont {Vergniory}}, \bibinfo {author}
  {\bibfnamefont {C.}~\bibnamefont {Felser}}, \bibinfo {author} {\bibfnamefont
  {R.}~\bibnamefont {Cava}}, \ and\ \bibinfo {author} {\bibfnamefont {B.~A.}\
  \bibnamefont {Bernevig}},\ }\href@noop {} {\bibfield  {journal} {\bibinfo
  {journal} {Science}\ }\textbf {\bibinfo {volume} {353}},\ \bibinfo {pages}
  {aaf5037} (\bibinfo {year} {2016})}\BibitemShut {NoStop}%
\bibitem [{\citenamefont {Tao}\ \emph {et~al.}(2013)\citenamefont {Tao},
  \citenamefont {Zhao}, \citenamefont {Chen}, \citenamefont {Li}, \citenamefont
  {Li}, \citenamefont {Ma}, \citenamefont {Li}, \citenamefont {Cui},
  \citenamefont {Zhu},\ and\ \citenamefont {Wang}}]{tao2013enhanced}%
  \BibitemOpen
  \bibfield  {author} {\bibinfo {author} {\bibfnamefont {Q.}~\bibnamefont
  {Tao}}, \bibinfo {author} {\bibfnamefont {X.}~\bibnamefont {Zhao}}, \bibinfo
  {author} {\bibfnamefont {Y.}~\bibnamefont {Chen}}, \bibinfo {author}
  {\bibfnamefont {J.}~\bibnamefont {Li}}, \bibinfo {author} {\bibfnamefont
  {Q.}~\bibnamefont {Li}}, \bibinfo {author} {\bibfnamefont {Y.}~\bibnamefont
  {Ma}}, \bibinfo {author} {\bibfnamefont {J.}~\bibnamefont {Li}}, \bibinfo
  {author} {\bibfnamefont {T.}~\bibnamefont {Cui}}, \bibinfo {author}
  {\bibfnamefont {P.}~\bibnamefont {Zhu}}, \ and\ \bibinfo {author}
  {\bibfnamefont {X.}~\bibnamefont {Wang}},\ }\href@noop {} {\bibfield
  {journal} {\bibinfo  {journal} {RSC Advances}\ }\textbf {\bibinfo {volume}
  {3}},\ \bibinfo {pages} {18317} (\bibinfo {year} {2013})}\BibitemShut
  {NoStop}%
\bibitem [{\citenamefont {Said}\ \emph {et~al.}(2011)\citenamefont {Said},
  \citenamefont {Sinn},\ and\ \citenamefont {Divan}}]{Said2011}%
  \BibitemOpen
  \bibfield  {author} {\bibinfo {author} {\bibfnamefont {A.~H.}\ \bibnamefont
  {Said}}, \bibinfo {author} {\bibfnamefont {H.}~\bibnamefont {Sinn}}, \ and\
  \bibinfo {author} {\bibfnamefont {R.}~\bibnamefont {Divan}},\ }\href
  {\doibase 10.1107/S0909049511001828} {\bibfield  {journal} {\bibinfo
  {journal} {Journal of Synchrotron Radiation}\ }\textbf {\bibinfo {volume}
  {18}},\ \bibinfo {pages} {492} (\bibinfo {year} {2011})}\BibitemShut
  {NoStop}%
\bibitem [{\citenamefont {Toellner}\ \emph {et~al.}(2011)\citenamefont
  {Toellner}, \citenamefont {Alatas},\ and\ \citenamefont
  {Said}}]{Toellner2011}%
  \BibitemOpen
  \bibfield  {author} {\bibinfo {author} {\bibfnamefont {T.~S.}\ \bibnamefont
  {Toellner}}, \bibinfo {author} {\bibfnamefont {A.}~\bibnamefont {Alatas}}, \
  and\ \bibinfo {author} {\bibfnamefont {A.~H.}\ \bibnamefont {Said}},\ }\href
  {\doibase 10.1107/S0909049511017535} {\bibfield  {journal} {\bibinfo
  {journal} {Journal of Synchrotron Radiation}\ }\textbf {\bibinfo {volume}
  {18}},\ \bibinfo {pages} {605} (\bibinfo {year} {2011})}\BibitemShut
  {NoStop}%
\bibitem [{\citenamefont {Kresse}\ and\ \citenamefont
  {Furthm{\"u}ller}(1996)}]{CAL_VASP}%
  \BibitemOpen
  \bibfield  {author} {\bibinfo {author} {\bibfnamefont {G.}~\bibnamefont
  {Kresse}}\ and\ \bibinfo {author} {\bibfnamefont {J.}~\bibnamefont
  {Furthm{\"u}ller}},\ }\href@noop {} {\bibfield  {journal} {\bibinfo
  {journal} {Physical review B}\ }\textbf {\bibinfo {volume} {54}},\ \bibinfo
  {pages} {11169} (\bibinfo {year} {1996})}\BibitemShut {NoStop}%
\bibitem [{\citenamefont {Gonze}\ and\ \citenamefont {Lee}(1997)}]{CAL_DFPT}%
  \BibitemOpen
  \bibfield  {author} {\bibinfo {author} {\bibfnamefont {X.}~\bibnamefont
  {Gonze}}\ and\ \bibinfo {author} {\bibfnamefont {C.}~\bibnamefont {Lee}},\
  }\href {\doibase 10.1103/PhysRevB.55.10355} {\bibfield  {journal} {\bibinfo
  {journal} {Phys. Rev. B}\ }\textbf {\bibinfo {volume} {55}},\ \bibinfo
  {pages} {10355} (\bibinfo {year} {1997})}\BibitemShut {NoStop}%
\bibitem [{\citenamefont {Perdew}\ \emph {et~al.}(1996)\citenamefont {Perdew},
  \citenamefont {Burke},\ and\ \citenamefont {Ernzerhof}}]{DFT_GGA}%
  \BibitemOpen
  \bibfield  {author} {\bibinfo {author} {\bibfnamefont {J.~P.}\ \bibnamefont
  {Perdew}}, \bibinfo {author} {\bibfnamefont {K.}~\bibnamefont {Burke}}, \
  and\ \bibinfo {author} {\bibfnamefont {M.}~\bibnamefont {Ernzerhof}},\
  }\href@noop {} {\bibfield  {journal} {\bibinfo  {journal} {Physical review
  letters}\ }\textbf {\bibinfo {volume} {77}},\ \bibinfo {pages} {3865}
  (\bibinfo {year} {1996})}\BibitemShut {NoStop}%
\bibitem [{\citenamefont {Liu}\ \emph {et~al.}(2014)\citenamefont {Liu},
  \citenamefont {Peng}, \citenamefont {Yin}, \citenamefont {Liu}, \citenamefont
  {Wang}, \citenamefont {Zhu}, \citenamefont {Wang}, \citenamefont {Liu},\ and\
  \citenamefont {He}}]{Liu2014}%
  \BibitemOpen
  \bibfield  {author} {\bibinfo {author} {\bibfnamefont {P.}~\bibnamefont
  {Liu}}, \bibinfo {author} {\bibfnamefont {F.}~\bibnamefont {Peng}}, \bibinfo
  {author} {\bibfnamefont {S.}~\bibnamefont {Yin}}, \bibinfo {author}
  {\bibfnamefont {F.}~\bibnamefont {Liu}}, \bibinfo {author} {\bibfnamefont
  {Q.}~\bibnamefont {Wang}}, \bibinfo {author} {\bibfnamefont {X.}~\bibnamefont
  {Zhu}}, \bibinfo {author} {\bibfnamefont {P.}~\bibnamefont {Wang}}, \bibinfo
  {author} {\bibfnamefont {J.}~\bibnamefont {Liu}}, \ and\ \bibinfo {author}
  {\bibfnamefont {D.}~\bibnamefont {He}},\ }\href {\doibase 10.1063/1.4872459}
  {\bibfield  {journal} {\bibinfo  {journal} {Journal of Applied Physics}\
  }\textbf {\bibinfo {volume} {115}},\ \bibinfo {pages} {163502} (\bibinfo
  {year} {2014})},\ \Eprint
  {http://arxiv.org/abs/https://doi.org/10.1063/1.4872459}
  {https://doi.org/10.1063/1.4872459} \BibitemShut {NoStop}%
\bibitem [{\citenamefont {Ding}\ \emph {et~al.}(2016)\citenamefont {Ding},
  \citenamefont {Shao}, \citenamefont {Zhang}, \citenamefont {Lu},
  \citenamefont {Ding}, \citenamefont {Ning},\ and\ \citenamefont
  {Huang}}]{ding2016crystal}%
  \BibitemOpen
  \bibfield  {author} {\bibinfo {author} {\bibfnamefont {L.-P.}\ \bibnamefont
  {Ding}}, \bibinfo {author} {\bibfnamefont {P.}~\bibnamefont {Shao}}, \bibinfo
  {author} {\bibfnamefont {F.-H.}\ \bibnamefont {Zhang}}, \bibinfo {author}
  {\bibfnamefont {C.}~\bibnamefont {Lu}}, \bibinfo {author} {\bibfnamefont
  {L.}~\bibnamefont {Ding}}, \bibinfo {author} {\bibfnamefont {S.~Y.}\
  \bibnamefont {Ning}}, \ and\ \bibinfo {author} {\bibfnamefont {X.~F.}\
  \bibnamefont {Huang}},\ }\href@noop {} {\bibfield  {journal} {\bibinfo
  {journal} {Inorganic chemistry}\ }\textbf {\bibinfo {volume} {55}},\ \bibinfo
  {pages} {7033} (\bibinfo {year} {2016})}\BibitemShut {NoStop}%
\bibitem [{sup()}]{supp}%
  \BibitemOpen
  \href@noop {} {\bibinfo  {journal} {See Supplemental Material at [URL will be
  inserted by publisher] for inelastic x-ray scattering cross section,
  structure factor calculations and spectra fittings}\ }\BibitemShut {NoStop}%
\bibitem [{\citenamefont {{Lopez Sancho}}\ \emph {et~al.}(1985)\citenamefont
  {{Lopez Sancho}}, \citenamefont {{Lopez Sancho}}, \citenamefont {{Sancho}},\
  and\ \citenamefont {{Rubio}}}]{CAL_GREEN1}%
  \BibitemOpen
\bibfield  {journal} {  }\bibfield  {author} {\bibinfo {author} {\bibfnamefont
  {M.~P.}\ \bibnamefont {{Lopez Sancho}}}, \bibinfo {author} {\bibfnamefont
  {J.~M.}\ \bibnamefont {{Lopez Sancho}}}, \bibinfo {author} {\bibfnamefont
  {J.~M.~L.}\ \bibnamefont {{Sancho}}}, \ and\ \bibinfo {author} {\bibfnamefont
  {J.}~\bibnamefont {{Rubio}}},\ }\href {\doibase 10.1088/0305-4608/15/4/009}
  {\bibfield  {journal} {\bibinfo  {journal} {Journal of Physics F Metal
  Physics}\ }\textbf {\bibinfo {volume} {15}},\ \bibinfo {pages} {851}
  (\bibinfo {year} {1985})}\BibitemShut {NoStop}%
\bibitem [{\citenamefont {Sancho}\ \emph {et~al.}(1984)\citenamefont {Sancho},
  \citenamefont {Sancho},\ and\ \citenamefont {Rubio}}]{CAL_GREEN2}%
  \BibitemOpen
  \bibfield  {author} {\bibinfo {author} {\bibfnamefont {M.~P.~L.}\
  \bibnamefont {Sancho}}, \bibinfo {author} {\bibfnamefont {J.~M.~L.}\
  \bibnamefont {Sancho}}, \ and\ \bibinfo {author} {\bibfnamefont
  {J.}~\bibnamefont {Rubio}},\ }\href
  {http://stacks.iop.org/0305-4608/14/i=5/a=016} {\bibfield  {journal}
  {\bibinfo  {journal} {Journal of Physics F: Metal Physics}\ }\textbf
  {\bibinfo {volume} {14}},\ \bibinfo {pages} {1205} (\bibinfo {year}
  {1984})}\BibitemShut {NoStop}%
\bibitem [{\citenamefont {Wu}\ \emph {et~al.}(2017)\citenamefont {Wu},
  \citenamefont {Zhang}, \citenamefont {Song}, \citenamefont {Troyer},\ and\
  \citenamefont {Soluyanov}}]{CAL_wanntools}%
  \BibitemOpen
  \bibfield  {author} {\bibinfo {author} {\bibfnamefont {Q.}~\bibnamefont
  {Wu}}, \bibinfo {author} {\bibfnamefont {S.}~\bibnamefont {Zhang}}, \bibinfo
  {author} {\bibfnamefont {H.-F.}\ \bibnamefont {Song}}, \bibinfo {author}
  {\bibfnamefont {M.}~\bibnamefont {Troyer}}, \ and\ \bibinfo {author}
  {\bibfnamefont {A.~A.}\ \bibnamefont {Soluyanov}},\ }\href
  {http://arxiv.org/abs/1703.07789} {\  (\bibinfo {year} {2017})},\ \Eprint
  {http://arxiv.org/abs/1703.07789} {arXiv:1703.07789} \BibitemShut {NoStop}%
\bibitem [{\citenamefont {Zhang}\ \emph {et~al.}(2019)\citenamefont {Zhang},
  \citenamefont {Jiang}, \citenamefont {Song}, \citenamefont {Huang},
  \citenamefont {He}, \citenamefont {Fang}, \citenamefont {Weng},\ and\
  \citenamefont {Fang}}]{zhang2019catalogue}%
  \BibitemOpen
  \bibfield  {author} {\bibinfo {author} {\bibfnamefont {T.}~\bibnamefont
  {Zhang}}, \bibinfo {author} {\bibfnamefont {Y.}~\bibnamefont {Jiang}},
  \bibinfo {author} {\bibfnamefont {Z.}~\bibnamefont {Song}}, \bibinfo {author}
  {\bibfnamefont {H.}~\bibnamefont {Huang}}, \bibinfo {author} {\bibfnamefont
  {Y.}~\bibnamefont {He}}, \bibinfo {author} {\bibfnamefont {Z.}~\bibnamefont
  {Fang}}, \bibinfo {author} {\bibfnamefont {H.}~\bibnamefont {Weng}}, \ and\
  \bibinfo {author} {\bibfnamefont {C.}~\bibnamefont {Fang}},\ }\href@noop {}
  {\bibfield  {journal} {\bibinfo  {journal} {Nature}\ }\textbf {\bibinfo
  {volume} {566}},\ \bibinfo {pages} {475} (\bibinfo {year}
  {2019})}\BibitemShut {NoStop}%
\bibitem [{\citenamefont {Vergniory}\ \emph {et~al.}(2019)\citenamefont
  {Vergniory}, \citenamefont {Elcoro}, \citenamefont {Felser}, \citenamefont
  {Regnault}, \citenamefont {Bernevig},\ and\ \citenamefont
  {Wang}}]{vergniory2019complete}%
  \BibitemOpen
  \bibfield  {author} {\bibinfo {author} {\bibfnamefont {M.}~\bibnamefont
  {Vergniory}}, \bibinfo {author} {\bibfnamefont {L.}~\bibnamefont {Elcoro}},
  \bibinfo {author} {\bibfnamefont {C.}~\bibnamefont {Felser}}, \bibinfo
  {author} {\bibfnamefont {N.}~\bibnamefont {Regnault}}, \bibinfo {author}
  {\bibfnamefont {B.~A.}\ \bibnamefont {Bernevig}}, \ and\ \bibinfo {author}
  {\bibfnamefont {Z.}~\bibnamefont {Wang}},\ }\href@noop {} {\bibfield
  {journal} {\bibinfo  {journal} {Nature}\ }\textbf {\bibinfo {volume} {566}},\
  \bibinfo {pages} {480} (\bibinfo {year} {2019})}\BibitemShut {NoStop}%
\bibitem [{\citenamefont {Tang}\ \emph {et~al.}(2019)\citenamefont {Tang},
  \citenamefont {Po}, \citenamefont {Vishwanath},\ and\ \citenamefont
  {Wan}}]{tang2019comprehensive}%
  \BibitemOpen
  \bibfield  {author} {\bibinfo {author} {\bibfnamefont {F.}~\bibnamefont
  {Tang}}, \bibinfo {author} {\bibfnamefont {H.~C.}\ \bibnamefont {Po}},
  \bibinfo {author} {\bibfnamefont {A.}~\bibnamefont {Vishwanath}}, \ and\
  \bibinfo {author} {\bibfnamefont {X.}~\bibnamefont {Wan}},\ }\href@noop {}
  {\bibfield  {journal} {\bibinfo  {journal} {Nature}\ }\textbf {\bibinfo
  {volume} {566}},\ \bibinfo {pages} {486} (\bibinfo {year}
  {2019})}\BibitemShut {NoStop}%
\bibitem [{\citenamefont {Baron}(2015)}]{Baron2015}%
  \BibitemOpen
  \bibfield  {author} {\bibinfo {author} {\bibfnamefont {A.~Q.~R.}\
  \bibnamefont {Baron}},\ }\enquote {\bibinfo {title} {High-resolution
  inelastic x-ray scattering i: Context, spectrometers, samples, and
  superconductors},}\ in\ \href {\doibase 10.1007/978-3-319-04507-8_41-1}
  {\emph {\bibinfo {booktitle} {Synchrotron Light Sources and Free-Electron
  Lasers: Accelerator Physics, Instrumentation and Science Applications}}},\
  \bibinfo {editor} {edited by\ \bibinfo {editor} {\bibfnamefont
  {E.}~\bibnamefont {Jaeschke}}, \bibinfo {editor} {\bibfnamefont
  {S.}~\bibnamefont {Khan}}, \bibinfo {editor} {\bibfnamefont {J.~R.}\
  \bibnamefont {Schneider}}, \ and\ \bibinfo {editor} {\bibfnamefont {J.~B.}\
  \bibnamefont {Hastings}}}\ (\bibinfo  {publisher} {Springer International
  Publishing},\ \bibinfo {year} {2015})\ pp.\ \bibinfo {pages} {1--68},\
  \Eprint {http://arxiv.org/abs/arXiv:1504.01098v5} {arXiv:1504.01098v5}
  \BibitemShut {NoStop}%
\bibitem [{\citenamefont {Miao}\ \emph
  {et~al.}(2018{\natexlab{b}})\citenamefont {Miao}, \citenamefont {Ishikawa},
  \citenamefont {Heid}, \citenamefont {Le~Tacon}, \citenamefont {Fabbris},
  \citenamefont {Meyers}, \citenamefont {Gu}, \citenamefont {Baron},\ and\
  \citenamefont {Dean}}]{Miao2018}%
  \BibitemOpen
  \bibfield  {author} {\bibinfo {author} {\bibfnamefont {H.}~\bibnamefont
  {Miao}}, \bibinfo {author} {\bibfnamefont {D.}~\bibnamefont {Ishikawa}},
  \bibinfo {author} {\bibfnamefont {R.}~\bibnamefont {Heid}}, \bibinfo {author}
  {\bibfnamefont {M.}~\bibnamefont {Le~Tacon}}, \bibinfo {author}
  {\bibfnamefont {G.}~\bibnamefont {Fabbris}}, \bibinfo {author} {\bibfnamefont
  {D.}~\bibnamefont {Meyers}}, \bibinfo {author} {\bibfnamefont {G.~D.}\
  \bibnamefont {Gu}}, \bibinfo {author} {\bibfnamefont {A.~Q.~R.}\ \bibnamefont
  {Baron}}, \ and\ \bibinfo {author} {\bibfnamefont {M.~P.~M.}\ \bibnamefont
  {Dean}},\ }\href {\doibase 10.1103/PhysRevX.8.011008} {\bibfield  {journal}
  {\bibinfo  {journal} {Phys. Rev. X}\ }\textbf {\bibinfo {volume} {8}},\
  \bibinfo {pages} {011008} (\bibinfo {year} {2018}{\natexlab{b}})}\BibitemShut
  {NoStop}%
\bibitem [{\citenamefont {Valla}\ \emph {et~al.}(1999)\citenamefont {Valla},
  \citenamefont {Fedorov}, \citenamefont {Johnson},\ and\ \citenamefont
  {Hulbert}}]{Valla1999}%
  \BibitemOpen
  \bibfield  {author} {\bibinfo {author} {\bibfnamefont {T.}~\bibnamefont
  {Valla}}, \bibinfo {author} {\bibfnamefont {A.~V.}\ \bibnamefont {Fedorov}},
  \bibinfo {author} {\bibfnamefont {P.~D.}\ \bibnamefont {Johnson}}, \ and\
  \bibinfo {author} {\bibfnamefont {S.~L.}\ \bibnamefont {Hulbert}},\ }\href
  {\doibase 10.1103/PhysRevLett.83.2085} {\bibfield  {journal} {\bibinfo
  {journal} {Phys. Rev. Lett.}\ }\textbf {\bibinfo {volume} {83}},\ \bibinfo
  {pages} {2085} (\bibinfo {year} {1999})}\BibitemShut {NoStop}%
\bibitem [{\citenamefont {Tamt$\ddot{o}$gl}\ \emph {et~al.}(2018)\citenamefont
  {Tamt$\ddot{o}$gl}, \citenamefont {Campi}, \citenamefont {Bremholm},
  \citenamefont {Hedegaard}, \citenamefont {Iversen}, \citenamefont {Bianchi},
  \citenamefont {Hofmann}, \citenamefont {Marzari}, \citenamefont {Benedek},
  \citenamefont {Ellis},\ and\ \citenamefont {Allison}}]{Tamtogl2018}%
  \BibitemOpen
  \bibfield  {author} {\bibinfo {author} {\bibfnamefont {A.}~\bibnamefont
  {Tamt$\ddot{o}$gl}}, \bibinfo {author} {\bibfnamefont {D.}~\bibnamefont
  {Campi}}, \bibinfo {author} {\bibfnamefont {M.}~\bibnamefont {Bremholm}},
  \bibinfo {author} {\bibfnamefont {E.~M.~J.}\ \bibnamefont {Hedegaard}},
  \bibinfo {author} {\bibfnamefont {B.~B.}\ \bibnamefont {Iversen}}, \bibinfo
  {author} {\bibfnamefont {M.}~\bibnamefont {Bianchi}}, \bibinfo {author}
  {\bibfnamefont {P.}~\bibnamefont {Hofmann}}, \bibinfo {author} {\bibfnamefont
  {N.}~\bibnamefont {Marzari}}, \bibinfo {author} {\bibfnamefont
  {G.}~\bibnamefont {Benedek}}, \bibinfo {author} {\bibfnamefont
  {J.}~\bibnamefont {Ellis}}, \ and\ \bibinfo {author} {\bibfnamefont
  {W.}~\bibnamefont {Allison}},\ }\href {\doibase 10.1039/C8NR03102A}
  {\bibfield  {journal} {\bibinfo  {journal} {Nanoscale}\ }\textbf {\bibinfo
  {volume} {10}},\ \bibinfo {pages} {14627} (\bibinfo {year}
  {2018})}\BibitemShut {NoStop}%
\end{thebibliography}%

\end{document}